\documentclass[xcolor=dvipsnames,a4,floatfix,amssymb,twocolumn]{revtex4}

\usepackage[dvipsnames]{xcolor}
\usepackage{amstext}
\usepackage{amsmath}
\usepackage{latexsym}

\usepackage[english]{babel}
\usepackage[latin1]{inputenc}

\usepackage{times}
\usepackage[T1]{fontenc}

\usepackage{graphics}
\usepackage{graphicx}

\usepackage{setspace}

\usepackage{textpos}

\newcommand{\be}{\begin{equation}}
\newcommand{\ee}[1]{\label{#1} \end{equation}}
\newcommand{\ba}{\begin{eqnarray}}
\newcommand{\ea}[1]{\label{#1} \end{eqnarray}}
\newcommand{\nl}{\nonumber \\}

\newcommand{\dt}[2]{ {{\textrm d} #1}/{{\textrm d} #2}}

\newcommand{\exv}[1]{{\, \left\langle {#1} \right\rangle \, }}

\definecolor{Pergamen}{RGB}{235,225,200}
\definecolor{LightGray}{RGB}{235,235,230}
\definecolor{PaleBlue}{RGB}{190,210,255}
\definecolor{DarkGreen}{RGB}{0,80,20}
\definecolor{SoftRed}{RGB}{255,220,170}

\usepackage[english]{babel}
\usepackage[latin1]{inputenc}

\usepackage{times}
\usepackage[T1]{fontenc}

\usepackage{graphics}
\usepackage{graphicx}
\usepackage{fancybox}

\usepackage{setspace}

\usepackage{textpos}

\begin{document}
\title[Reservoir Fluctuations] 
{Statistical Power-Law Spectra due to Reservoir Fluctuations 
}

\author{T.S.~Bir\'o,  G.G.~Barnaf\"oldi,  P.~V\'an and K.~\"Urm\"ossy}
\affiliation{
  Heavy Ion Research Group\\
  MTA 
  Wigner Research Centre for Physics, Budapest 
}

\date{\today}


\begin{abstract}


LHC ALICE data are interpreted in terms of statistical
power-law tailed $p_T$ spectra. As explanation
we derive such statistical distributions for
particular particle number fluctuation patterns in
a finite heat bath exactly, and for general
thermodynamical systems in the subleading canonical
expansion approximately. Our general
result, $q=1-1/C+\Delta T^2/T^2$, demonstrates how
the heat capacity and the temperature fluctuation effects
compete, and cancel only in the standard Gaussian approximation.

\end{abstract}






\maketitle



\section{Introduction}

Power-law tailed distributions occur in Nature numerous.
The idea of a statistical -- thermodynamical origin
of these emerged already decades ago \cite{TsallisOrigPaper,TsallisBook}.
We have, however, long missed a ''naturalness''
argument connecting the basic principles of classical
thermodynamics to the use of non-extensive entropy
formulas by deriving canonical distributions of the one-particle
energy. Although the observation has been made that the
Tsallis and R\'enyi entropy formulas both lead to the 
cut power-law canonical distribution, and their use requires
a constant heat capacity reservoir \cite{Almeida}, the $q>1$ power-laws
-- featuring a negative power of a quantity larger than one --
still seem unnatural.

In recent studies of ideal gases \cite{BiroPHYSA2013,BAGCI,CAMPISI}
we investigated
energy fluctuations in a subsystem \,--\, reservoir couple.
They lead to Tsallis distribution with $q=1-1/C$
for ideal gas reservoirs, with $C$ being the heat capacity 
of the total system. 

Moreover, particle number fluctuations in the reservoir,
either achieved naturally in a huge, inhomogenous
heat bath or artificially by averaging the statistics
over repeated events in high-energy experiments,
lead to further effects \cite{Wilk,Wilk2,Begun1,Begun2}. 
We review in this paper how ideal fermionic and bosonic
reservoirs, with binomial (BD) and  negative binomial (NBD)
distributions of the particle number, lead exactly to Tsallis power-law
behavior with the parameters $T=E/\langle n\rangle$ and 
$q=\langle n(n-1) \rangle / \langle n\rangle^2$,
when the microcanonical ideal gas statistical factor, 
$(1-\omega/E)^n$ in one dimension for massless partons,
\footnote{This is $\exp(S(E-\omega)-S(E))$, 
the complement phase-space factor
for ideal gases. Since each exponential grows like
$x^n$, their ratio delivers the formula.}
is averaged over one of these distributions.
The above $q$, named as second factorial moment, $F_2$, 
was determined with respect to canonical suppression in Refs.~\cite{KOCH,BEGUN}.
For the binomial distribution one gets $q=1-1/k$, for
the negative binomial $q=1+1/(k+1)$.

We demonstrate by fits to recent ALICE data taken in LHC experiment~\cite{ALICE2013}
that in the $p_T$-distribution of charged hadrons (dominated
by pions) two Tsallis distributions emerge for the one-particle energy
in a moving system, $\omega=\gamma(m_T-vp_T)$
(with $\gamma=1/\sqrt{1-v^2}$ being the Lorentz factor and
$v$ a radial blast wave velocity, $m_T=\sqrt{m^2+p_T^2}\approx |p_T|$
the so called transverse mass). The softer parts, below $p_T \approx 4$ GeV/c,
show a dependence on the participant number as expected from
statistical considerations: bigger systems come closer to the
Boltzmann\,--\,Gibbs prediction.

Our theoretical results on $q$ and $T$ expressed by the mean
multiplicity and its variance in the reservoir for BD and NBD
distributions also can be viewed as an approximation
for arbitrary particle number distributions in the reservoir
up to subleading (second) order in the canonical expansion 
$\omega \ll E$.  For non-ideal systems the general expansion 
up to second order delivers $q=1-1/C+\Delta T^2/T^2$, a combined
result with the heat capacity and the variance of the 
temperature of finite heat bath. These quantities seem to
act against each other. Here the variance of the temperature
is meant for the estimator $1/S'(E)$ of the thermodynamical 
temperature, the latter defined by $1/T = \langle S'(E) \rangle$.
This way in the Gaussian approximation $\Delta T/T = 1/\sqrt{C}$
we regain $q=1$ and verify the Boltzmann\,--\,Gibbs statistical factor.
Part of this result has been derived and promoted by G.~Wilk
and Z.~Wlodarczyk ($q=1+\Delta T^2/T^2$) in recent years
\cite{Wilk4,Wilk5,Wilk6}. Instead of temperature fluctuations
reservoir volume and particle number fluctuations were considered in 
recent publications~\cite{Begun1,Begun2,Gorenstein,Gorenstein2}.


\section{$p_T$ spectra at the LHC}

In high-energy physics the power-law tail in $p_T$ spectra is traditionally
fitted by cut power-laws, $(1+ap_T)^{-b}$, conjectured to stem from
the behavior of hadronization matrix elements.
As a matter of fact, a statistical model also can be applied to the
fragmentation functions which describe the yield of hadrons stemming from
high-energy particle jets~\cite{UrmossyPLB2011,UrmossyPLB2012}.
The real unknown is the soft part, with low $p_T$ momenta; here thermal models
are more fashionable. 

It is therefore an intriguing question to decide whether there is a soft
power-law, which can be naturally described and understood only by statistical phase-space
considerations. The idea of a cut power-law as a thermal distribution,
a characteristic consequence also from non-extensive thermodynamics,
has been pursued by us since several years \cite{BiroPRL2005,BiroBook2011,BiroJPG2010}.
It is now for the first time that particle spectra over a wide $p_T$ range
are presented differentially for centrality classes \cite{ALICE2013}; such
a presentation may inform about the
multiplicity dependence of a heat reservoir in terms of thermal models.

\begin{figure}

	\includegraphics[height=0.32\textheight]{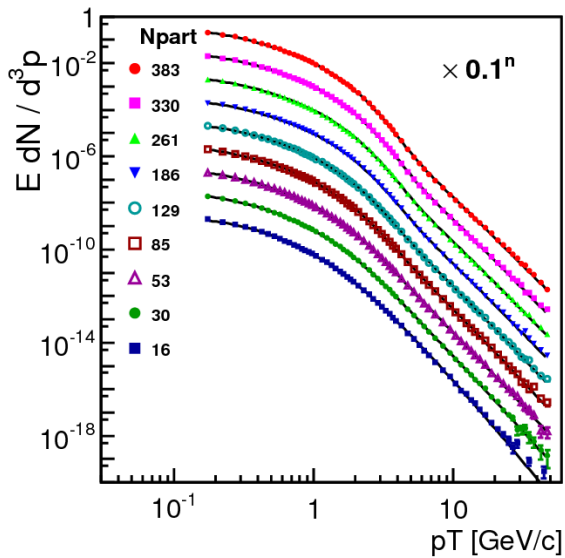}

\caption{\label{Fig1}
  $(1+ap_T)^{-b}$ fits to ALICE data on charged hadron $p_T$ spectra in PbPb 
  collisions\cite{ALICE2013} at LHC show two power-laws.
  Fit parameters as function of $N_{{\rm part}}$ are shown in Fig.~\ref{Fig2}.
}
\end{figure}

In Fig.~\ref{Fig1} we display our fits to $p_T$ spectra of charged hadrons
in centrality classes.
A break in the spectra is pronounced at high centralities
(large participant numbers, $N_{{\rm part}}$), which must be positively correlated with
the particle number in the fireball where the hadrons were born.
Our fits have the lowest $\chi^2$ by making the 
soft\,--\,hard change around $p_T\approx 4$ GeV/c for all
centrality classes, therefore we think it is justified to talk about
soft and hard power-laws separately.



\begin{figure}

\begin{center}

\includegraphics[height=0.32\textheight]{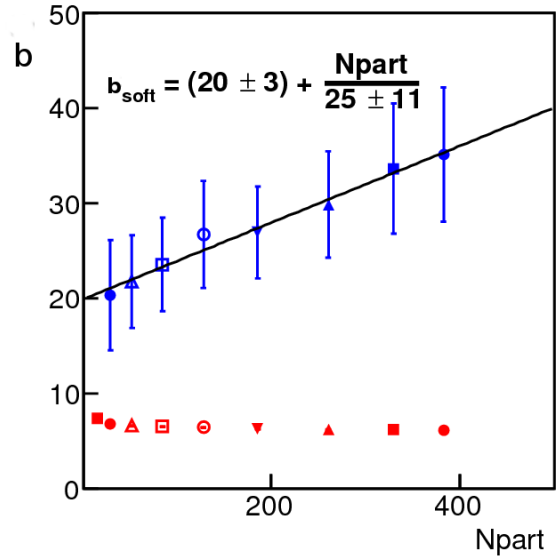}

\end{center}

\caption{\label{Fig2}
Powers in  the power law, $b=1/(q-1)$, follow a statistical trend for the soft 
spectra (upper symbols), while remain nearly constant for the hard spectra (lower symbols).
The results belong to the participant numbers, $N_{{\rm part}}$, seen in the legend
of Fig.~\ref{Fig1}.
}	
\end{figure}

The fit parameter $b$, connected to the parameter $q$ in Tsallis distribution as $b=1/(q-1)$,
is plotted against $N_{{\rm part}}$ in Fig.~\ref{Fig2}. 
The soft part shows a clear rising of the power $b$ with
$N_{{\rm part}}$, very characteristic to a statistical -- thermal origin
of a power-law.
Contrary to this is the behavior of the hard spectra: the fitted power stays constant
irrespective to the centrality, conjectured to vary with the size of the thermal bath.
This is 'naturally' expected from QCD.

\section{Temperature and Energy Fluctuations}


In this Section we turn to the theory of statistical power-law tailed
distributions as canonical distributions in a thermal system connected
to a heat reservoir with finite heat capacity. By fluctuation of
temperature we mean the fluctuation of the estimator $1/S'(E)$ due
to fluctuations of the energy $E$ in the reservoir. We are interested
in the observable distribution of the one-particle energy, $\omega \ll E$,
in the canonical limit. 

Traditionally such thermodynamical fluctuations are treated in the
Gaussian approximation. Based on the fundamental thermodynamic
uncertainty relation, $\Delta E \cdot \Delta \beta = 1$ 
with $\beta=S'(E)$, it is easy to derive the characteristic scaled
fluctuation of the temperature ~\cite{Uffink1,Lavenda,Uffink2}.
With any well peaked distribution of a random variable, $x$,
the expectation value $a=\exv{x}$ is near to the value where the peak occurs. 
As a consequence the variance of any function, $f(x)$ in this approximation
is related to the original variance by a Jacobi determinant: $\Delta f = |f'(a)| \Delta x$.
Now we consider both $E$ and $\beta$ as functions of the temperature, $T$.
We obtain $\Delta E = |C| \Delta T$ with $C=\dt{E}{T}$ being the definition of heat capacity,
and $\Delta \beta = \Delta T /T^2$. Combining these two results one arrives at
the classical formula $\Delta T/T = 1/\sqrt{|C|}$.
The heat capacity $C$ is proportional to the heat bath size (volume, number of degrees of freedom)
for large extensive systems.

There are, however, some deficiences in the Gaussian
approximation. A Gauss distribution of $\beta$,  given as
$w(\beta) \propto \exp\left({-{C}(T\beta-1)^2/2}\right)$,
allows for a finite probability for negative temperatures, and 
-- even worse -- its characteristic function, 
$\exv{e^{-\beta\omega}} = \exp\left({-\omega/T+\omega^2/2CT^2}\right)$
is not integrable in $\omega$.


The next theoretical question is how to improve the canonical scheme 
beyond the Gauss approximation.
We start our discussion with ideal gases.
The one-particle energy, $\omega$, out of total energy, $E$, 
is distributed according to a statistical weight 
factor $(1-\omega/E)^n$~\footnote{This statistical weight factor is to 
be multiplied with a one-particle phase-space factor, $\rho(\omega)$, 
which however does not depend on $n$ and $E$.}. 
The idea of superstatistics in general considers
a distribution for the reservoir parameters $n$ and $E$ \cite{SuperStat,Beck2}.
In high-energy experiments $E$ is typically controlled by the accelerator
and does not vary much. However, $n$, the number of particles in the produced
fireball scatters appreciably, which can be uncovered via the 
event-by-event detection of the spectra in $\omega$,
as suggested in~\cite{VHMPID}.

In ideal reservoirs  $n$ particles are distributed
among $k$ phase-space cells: fermions $\binom{k}{n}$, bosons $\binom{n+k}{n}$  ways.
The binomial and negative binomial distributions 
can be derived by considering a subspace $(n,k)$ out of $(N,K)$
in the limit \hbox{$K\rightarrow\infty$} and $N\rightarrow\infty$ while $f=N/K$ is fixed.
\be
F_{n,k}(f) := \lim_{K\rightarrow\infty}\limits 
\frac{\binom{k}{n}\binom{K-k}{N-n}}{\binom{K}{N}}
= \binom{k}{n} f^n (1-f)^{k-n}.
\ee{BD_DEF}
\ba
B_{n,k}(f) := \lim_{K\rightarrow\infty}\limits 
\frac{\binom{n+k}{n}\binom{N-n+K-k}{N-n}}{\binom{N+K+1}{N}}
\nl
= \binom{n+k}{n} f^n (1+f)^{-n-k-1}.
\ea{NBD_DEF}
These distributions are normalized based on the binomial expansion of
$(a+b)^k$ and $(b-a)^{-k-1}$, respectively.

Assuming a typical fireball in high-energy experiments,
$E$ is fixed and $n$ fluctuates according to NBD. The ideal gas
statistical weight factor, describing the complement phase-space
for reservoir configurations, becomes~\footnote{One has to note that in fact zero particles
in the reservoir are irrealistic. Partial sums starting not at $n=0$ can be obtained
by derivations with respect to $\omega$.}
\be
 \sum_{n=0}^{\infty}\limits \left(1-\frac{\omega}{E} \right)^n B_{n,k}(f) =
 \left(1+ f\frac{\omega}{E} \right)^{-k-1}.
\ee{NBD_AVER}
Note that $\exv{n}=(k+1)f$ for NBD. Then with $T=E/\exv{n}$ and $q=1+1/(k+1)$
we get
\be
\left(1 + (q-1) \frac{\omega}{T} \right)^{-\frac{1}{q-1}}.
\ee{TSALLIS_DIST}
This is {\em exactly} the statistical weight factor
which provides the $q>1$ Tsallis\,--\,Pareto 
distribution\footnote{Unfortunately different conventions are in use for the parameter
$q$, some papers in fact apply $2-q$ at the same place.}.
Similarly in a fermionic reservoir $E$ is fixed, $n$ is distributed according to BD.
We obtain
 \be
 \sum_{n=0}^{\infty}\limits \left(1-\frac{\omega}{E} \right)^n F_{n,k}(f) =
 \left(1- f\frac{\omega}{E} \right)^{k}.
 \ee{BD_AVER}
Note that $\exv{n}=kf$ for BD. Then with $T=E/\exv{n}$ and $q=1-1/k$
we again get a Tsallis\,--\,Pareto distribution, but now with $q < 1$.
In the $k \gg n$ limit (low occupancy in phase-space) the particle distribution
in the reservoir becomes Poissonian in both cases.
The result is {exactly} the Boltzmann\,--\,Gibbs weight factor with $T=E/\exv{n}$:
\be
\sum_{n=0}^{\infty}\limits \left(1 - \frac{\omega}{E}\right)^n \frac{\exv{n}^n}{n!} e^{-\exv{n}} =
e^{-\exv{n} \: \omega/E}.
\ee{POIS_AVER}
We note that NBD distributions are observed experimentally, a nice analysis
of heavy ion data are given by the PHENIX group~\cite{PHENIXNBD}.
In all the three above cases
\be
T = \frac{E}{\exv{n}}   \quad \textrm{and} \quad q = \frac{\exv{n(n-1)}}{\exv{n}^2}.
\ee{ALLCASES}
Now we turn to the ideal statistical weight factor with general
finite reservoir fluctuations.
In the canonical approach we expand for small $\omega \ll E$
and view the Tsallis\,--\,Pareto distribution as an approximation:
\be
\left(1+(q-1)\frac{\omega}{T} \right)^{-\frac{1}{q-1}} =
1-\frac{\omega}{T} + q \frac{\omega^2}{2T^2} - \ldots
\ee{TSALLIS_EXPAND}
on the one hand and
\be
\exv{\left(1-\frac{\omega}{E}\right)^n} =
1 - \exv{n}\frac{\omega}{E} + \exv{n(n-1)}\frac{\omega^2}{2E^2} - \ldots
\ee{RESERV_EXPAND}
on the other hand.
To match up to subleading canonical order, it follows in general:
\be
T = \frac{E}{\exv{n}}  \quad \textrm{and} \quad q = \frac{\exv{n(n-1)}}{\exv{n}^2}.
\ee{TSALLIS_PARAMS_SECOND_ORDER}


Finally we consider a general system with general reservoir fluctuations.
Expanding for small $\omega \ll E$
\ba
\exv{e^{S(E-\omega)-S(E)}}_{\omega\ll E} = \exv{e^{-\omega S^{\prime}(E)+\omega^2 S^{\prime\prime}(E)/2 - \ldots}}
\nl
= 1 - \omega \exv{S^{\prime}(E)} + \frac{\omega^2}{2} \exv{S^{\prime}(E)^2+S^{\prime\prime}(E)} - \ldots
\ea{COMPLEMENT_PHASE_SPACE}
Compare this with the expansion eq.(\ref{TSALLIS_EXPAND}) of the Tsallis distribution:
In the view of the above we interpret the parameters as
\be
\frac{1}{T} = \exv{S^{\prime}(E)}, \qquad 
q = \frac{\exv{S^{\prime}(E)^2 + S^{\prime\prime}(E)}}{\exv{S^{\prime}(E)}^2}.
\ee{INTERPRET}
Here  $\exv{S^{\prime\prime}(E)}=-1/CT^2$ 
follows from the definition of the heat capacity of the reservoir, $1/C = \dt{T}{E}$.
Summarizing these results we understand that the parameter $q$
has opposite sign contributions from $\exv{S^{\prime \: 2}}-\exv{S^{\prime}}^2$ 
and from $\exv{S^{\prime\prime}}$. In general
\be
 q = 1 + \frac{\Delta T^2}{T^2} - \frac{1}{C}. 
\ee{IMPORTANT_RESULT}
to subleading canonical order. With this formula
$q>1$ and $q<1$ are both possible and  for temperature fluctuations 
with Gaussian variance, $\Delta T/T = 1/\sqrt{C}$, one has $q=1$.

%

\vspace{3mm}
{In summary} we studied the mechanism behind the occurence of statistical power-law
distributions in high-energy particle creation, exploiting the role of
reservoir fluctuations. First we demonstrated that ALICE charged hadron
$p_T$-spectra feature soft, statistical Tsallis-distributions besides
the traditionally known hard QCD based power-law. Observing that the
power, $b=1/(q-1)$, in our fits increases with increasing participant number,
$N_{{\rm part}}$, we concluded that bigger fireballs come closer to
the conventional thermal model exponential distribution ($q \to 1$).

Seeking for a theoretical explanation we analyzed scenarios with
ideal gas reservoirs formed at fix total energy, $E$, but fluctuating
particle number $n$. We have clearly concluded that $q > 1$ power-laws
occur with the NBD distribution ($q=1+1/(k+1)$) and for the Poissonian
distribution of $n$ exactly the Boltzmann\,--\,Gibbs exponential formula
($q=1$) follows.

For a general fluctuation pattern of $n$ the Tsallis\,--\,Pareto
form is only an approximation, but it goes beyond the
traditional exponential. In general $T=E/\exv{n}$ and $q=\exv{n(n-1)}/\exv{n}^2$.
This interprets the parameter $q$ as the scaled second factorial moment
for ideal gas reservoirs with arbitrary particle number fluctuations.

Finally for non-ideal reservoirs, described by an equation of state,
$S(E)$, we also derived the meaning of the Tsallis parameters $T$ and $q$
(cf. eq.\ref{INTERPRET}). This is one of the main novel results of this paper.

Our formula demonstrates that in general heat capacity and temperature
variance effects compete with each other; at exact balance the
traditional Boltzmann\,--\,Gibbs thermodynamics is restored.
In this case the scaled fluctuations follow the traditional
inverse square root law.

As an outlook we shall consider that for cases when the entropy is
non-extensive the concept of reservoir has to be investigated in
more depth. First steps towards such an analysis included the construction
of a deformed entropy formula without, however, discussing fluctuations
in the resevoir~\cite{BiroEPJA2013}. 
The generalization of that procedure with reservoir fluctuations,
as discussed in this paper, will be presented in a forthcoming publication.
Since such an entropy concept also has to satisfy basic thermodynamical
requirements for a general equilibrium state, a deformed entropy formula
is not arbitrary~\cite{BiroPRE2011}.

\vspace{3mm}
{\bf Acknowledgement} ~ This work was supported by Hungarian
OTKA grants K81161, K104260, NK106119, and
NIH TET\_12\_CN-1-2012-0016. Author GGB also thanks
the J\'anos Bolyai Research Scholarship of the Hungarian
Academy of Sciences.


\end{document}